\documentclass{elsart}
\usepackage{amsfonts}
\usepackage{amsmath}
\usepackage{graphicx}

\begin{document}

\begin{frontmatter}

\title{Elasticity of strongly stretched  ssDNA}
\author{Alexei V. Tkachenko}

\address{Department of Physics and Michigan Center for Theoretical Physics, \\ University of Michigan, 450 Church Str.,  Ann Arbor MI 48109 }

\begin{abstract}
We present a simple model which describes elastic response of single-stranded DNA (ssDNA) to stretching, including  the  regime of very high  force (up to 1000 pN). ssDNA is modelled as a discreet  persistent chain,  whose ground state is a zigzag rather than a straight line configuration. This mimics the underlying molecular architecture and helps to explain the experimentally observed staturation of the stretching curve at very high force.   

\end{abstract}

\begin{keyword}
DNA, elasticity
\end{keyword}

\end{frontmatter}

\section{Introduction}

Over the recent years, impressive progress has been made in understanding of
the physics of DNA. To a large extent, this development was due to the
introduction of novel micromanipulation techniques suitable for study of the
mechanical response of a single molecule \cite{DNAstretch}\cite{ssDNAstretch}%
. The mechanical properties of double-stranded DNA (dsDNA) have been studied
in pioneering experiments of Smith et al \cite{DNAstretch}. These
experiments agree remarkably well with Worm-Like Chain (WLC) model which
describes DNA as an elastic rod, whose bending modulus is proportional to
chain persistence length $l_{p}$ ($l_{p}\approx 50nm$ for dsDNA, which is
the only adjustable parameter of the model) \cite{Marko}.

In the case of single-stranded DNA (ssDNA) and RNA molecules the choice of
an adequate elastic description remains an open problem. On the one hand,
ssDNA and RNA were traditionally described by Freely Joint Chain model
(FJC). Its extensible version has been originally used for fitting of the
early ssDNA stretching data \cite{ssDNAstretch}. On the other hand, the
chemical structure of ssDNA strongly support the picture with a finite
bending modulus, reminiscent of WLC model. However, being a continuous
model, the WLC description is unlikely to be valid in the regimes when
discrete nature of chemical bonds becomes relevant (e.g. for a sufficiently
high stretching force). To overcome these limitations, Discreet Persistent
Chian (DPC) model has been proposed for ssDNA by Storm and Nelson \cite%
{DPCnels}. Interestingly, WLC itself was introduced as a continuous limit of
a similar discrete model, originally proposed back in late 40s by Kratky and
Porod (KP) \cite{KP}. \ Authors of Ref. \cite{DPCnels} \ succeeded in
calculating the response of DPC to an arbitrary stretching force, by
combining the Transfer Matrix method with a variational procedure.
Nevertheless, the discreet model yields only a marginal improvement of the
fit of ssDNA force--extension curve, compared to WLC and FJC cases.

In order to improve the fitting, the extensible versions of the above models
are typically used, i.e. the linear deformations of the bonds are
introduced. However, the AFM experiments done in the limit of very high
force (up to 1000 pn) \cite{rief}, exhibit surprising saturation of the
end-to-end distance at certain value \textit{above} the chain counter
length. \ Any model which treat deformations of the ssDNA backbone in a
linear manner, is unable to capture this phenomenon. The experimental
behavior is however consistent with results of detailed \textit{ab-initio}
calculations \cite{abinitio}. In this paper, we propose a simple model
motivated by actual microscopic architecture of ssDNA. Without introduction
of extra fitting parameters, our "zigzag" model adequately describe both
low- and high-force regimes of ssDNA stretching.

\section{Zigzag model for ssDNA elasticity}

Discreet Persistent Chian (DPC), also known as Kratky and Porod (KP) model,
has the following Hamiltonian:

\begin{equation}
H=\sum\limits_{i}\left[ \frac{J}{2}\Theta _{i,i+1}^{2}-b\mathbf{f\cdot \hat{t%
}}_{i}\right]   \label{DPC}
\end{equation}%

Here $\mathbf{\hat{t}}_{i}$, is $i$-th bond orientation, $b$ is (fixed) bond
length, $\Theta _{i,i+1}$ is the angle between $i-th$ and $i+1$-th bonds
(i.e cos $\Theta _{i,i+1}=\mathbf{\hat{t}}_{i}\mathbf{\hat{t}}_{i+1}$), and $%
\mathbf{f}$ is external stretching force. \ Interestingly, two other common
models can be obtained from DPC \ Hamiltonian, by taking limits of zero bond
length (WLC), or zero coupling $J$ (FJC).

\begin{figure}[htbp]
\begin{center}
\includegraphics[height=2.5in,width=2in]{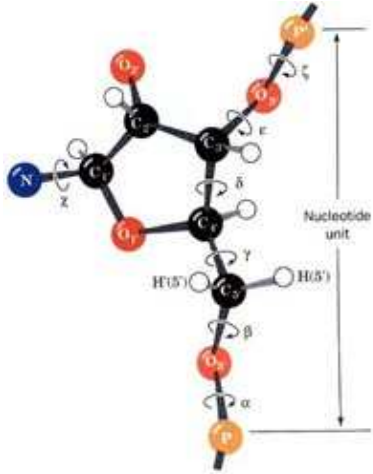}
\end{center}
\caption{Microscopic architecture of ssDNA.}
\label{torsion}
\end{figure}

At the atomistic level, the backbone conformation of ssDNA is often
parameterized by several torsional angles per nucleotide,
as shown in Figure \ref{torsion}. From this point of view, WLC and DPC models
describe small deviations from the ground state configuration. . However, at
the microscopical level, all the valent angles have certain preferred values
different from 180$^{0}$. The minimal generalization of DPC model which
mimics this observation is a chain whose ground state is zigzag (all-trans)
configuration: 
\begin{equation}
H=\sum\limits_{i}\left[ \frac{J}{2}\left( \Theta _{i,i+1}+\left( -1\right)
^{i}\Theta _{0}\cos \psi _{i,i+1}\right) ^{2}-b\mathbf{f\cdot \hat{t}}_{i}%
\right] .
\end{equation}%
Here $\psi _{i}$ is the torsional angle associated with rotation of $i-th$
bond. For simplicity, we assign the same preferred angle $\Theta _{0}$ to
each bond pair. Surprisingly, this minimal model results in a significant
improvement over the previous attempts to describe the experimentally
observed large force behavior of ssDNA.

\section{Mapping on DPC model}

In order to calculate the stretching curve of the "zigzag" chain, we note
that the model can be mapped back onto DPC model, with renormalized
parameters. Let us consider a chain consisting only of odd (or only even)
nodes of the zigzag (i.e. ...$-"i"-"i+2"-"i+4"-$...). The ground state of
this chain is a straight line, but the bond length (even without thermal
fluctuation) is a function of applied force: $b^{\ast }=2b\cos \theta ^{\ast
}$, where $\theta ^{\ast }\left( f\right) $ is the preferred bond
orientation angle $\theta _{i}$, with respect to the direction of $\mathbf{f}
$. In ground state configuration (i.e. zigzag) $\theta _{i}=\left( -1\right)
^{i}\theta ^{\ast }$ \ The value of $\theta ^{\ast }$ may be obtained by
minimizing the Hamiltonian, at given stretching force: 
\begin{equation}
H\left( \theta ^{\ast }\right) =\sum\limits_{i}\left[ \frac{J}{2}\left(
2\theta ^{\ast }-\Theta _{0}\right) ^{2}-fb\cos \theta ^{\ast }\right]
\end{equation}%
In the limit of small $\Theta _{0}$, we obtain::%
\begin{equation}
\theta ^{\ast }\left( f\right) =\left( \frac{bf}{2J}+1\right) ^{-1}\frac{%
\Theta _{0}}{2}
\end{equation}%
\begin{equation}
b^{\ast }\left( f\right) =2b\cos \Theta ^{\ast }\approx 2b\left( 1-\left( 
\frac{bf}{2J}+1\right) ^{-2}\frac{\Theta _{0}^{2}}{8}\right)
\end{equation}

It is also easy to demonstrate that the renormalized coupling in the final
DPC Hamiltonian will be $J^{\ast }=J/2$. The parameters of the renormalized
model are related to the experimentally observable persistence length as $%
l_{p}=b^{\ast }\left( 0\right) J^{\ast }/kT\approx Jb/kT$. We can now use
DPC model to calculate the stretching curve $z/L=x\left( f\right) $, and
include the effect of renormalization in the following manner:
\begin{equation}
\frac{z}{L}=\frac{b^{\ast }\left( f\right) }{b^{\ast }\left( 0\right) }%
x\left( f\right) \approx \left( 1+\left[ 1-\left( 1+\left( \frac{b}{l_{p}}%
\right) ^{2}\frac{fl_{p}}{2kT}\right) ^{-2}\right] \frac{\Theta _{0}^{2}}{8}%
\right) x\left( f\right)
\end{equation}%
Here $z$ and $L$ are end-to-end distance and the original chain counter
length, respectively. Since the discreetness \ of DPC is only marginally
important, one can relate $x$ to $f$ with a simple interpolation formula
proposed by Marko and Siggia for continuous WLC model:%
\begin{equation}
\frac{fl_{p}}{kT}=\frac{1}{4}\left( \frac{1}{\left( 1-x\right) ^{2}}%
-1\right) +x
\end{equation}

\begin{figure}[htbp]
\begin{center}
\includegraphics[height=3.5in,width=4.8in]{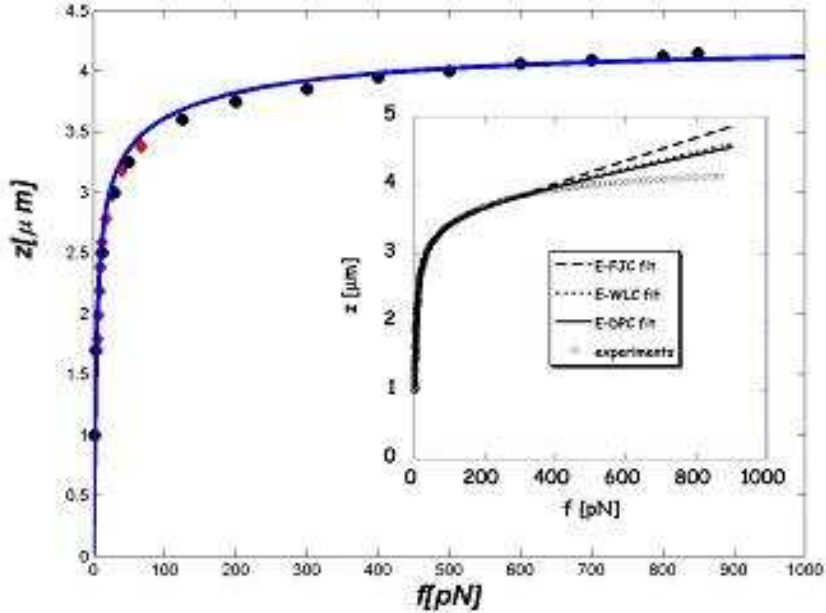}
\end{center}
\caption{Stretching behavior of ssDNA,
given by the zigzag model. Experimental points are from Refs \protect\cite%
{rief} (circles) and \protect\cite{ssDNAstretch}(diamonds). Insert: Fitting
of the experimental data with extensible FJC, WLC and DPC models 
\protect\cite{DPCnels}.}
\label{zigzag}
\end{figure}

The resulting stretching curve is shown in Figure \ref{zigzag}. It is
remarkable that the very good fit can be achieved with the same number of
free parameters, as in the earlier models: $b$(which is fixed at its
physical value of a typical chemincal bond length, $0.1nm$ ), ssDNA
persistence length $l_{p}\approx 0.85nm$ (extracted from the moderate force
stretching behavior). The only fitting in our model is angle $\Theta
_{0}/2\approx 25^{0}$ which essentially \ replaces the bond rigidity
commonly used in the extensible versions of classical models.

\section{Conclusions}

In conclusion, we proposed a new model for ssDNA elasticity. It differs from
the earlier ones by an explicit account for certain features of microscopic
architecture of the chain backbone. Qualitatively, our \ model attribute the
ultra-high-force stretching behavior to the perturbations to the valent
angles rather than deformations of the bonds themselves. This implication
can be directly checked by the first principle calculations.

\end{document}